\documentclass[aip,amsmath,amssymb,reprint,apl,floatfix]{revtex4-1}

\usepackage{graphicx}
\usepackage{dcolumn}
\usepackage{bm}
\usepackage{hyperref}
\usepackage{mathrsfs,amsmath} 

\usepackage{pgfpages}

\newcommand{\0}{{\lower 0.2ex\hbox{\kern 0.02em$\scriptstyle 0$}}} 

\def\zhat{\mbox{\boldmath${\hat{\mathrm{z}}}$}}

\def\DelphiE{\Delta\phi_{\lower2pt\hbox{$\scriptstyle E$}}}
\def\DelphiM{\Delta\phi_{\lower1pt\hbox{$\scriptstyle M$}}}

\def\fm{f_{\lower 0.4ex\hbox{\begin{scriptsize}M\end{scriptsize}}}} 
\def\fs{f_{\lower 0.4ex\hbox{\begin{scriptsize}S\end{scriptsize}}}} 
\def\ft{f_{\lower 0.4ex\hbox{\begin{scriptsize}T\end{scriptsize}}}} 
\def\ni{\noindent} 
\def\phim{\phi_{\lower 0.4ex\hbox{\begin{tiny}M\end{tiny}}}} 
\def\phis{\phi_{\lower 0.4ex\hbox{\begin{tiny}S\end{tiny}}}} 
\def\phiso{\phi_{\lower 0.4ex\hbox{\begin{tiny}S0\end{tiny}}}} 
\def\phitau{\phi_{\lower 0.4ex\hbox{\begin{footnotesize}$\tau$\end{footnotesize}}}}
\def\phit{\phi_{\lower 0.4ex\hbox{\begin{tiny}T\end{tiny}}}}
\def\phitauo{\phi_{\lower 0.4ex\hbox{\begin{footnotesize}$\tau$\end{footnotesize}\begin{scriptsize}0\end{scriptsize}}}} 
\def\Sch{Schr\"odinger} 
\def\subone{{\lower 0.3ex\hbox{$\scriptscriptstyle 1$}}} 
\def\subtwo{{\lower 0.3ex\hbox{$\scriptscriptstyle 2$}}} 
\def\subl{{\lower 0.3ex\hbox{$\scriptscriptstyle l$}}} 

\makeatletter
\newcommand{\mathleft}{\@fleqntrue\@mathmargin0pt}
\newcommand{\mathcenter}{\@fleqnfalse}
\makeatother

\begin{document}


\title{Space-time imaging, magnification and time reversal of matter waves} 



\author{Brian H.\ Kolner}
\email[]{bhkolner@ucdavis.edu}
\affiliation{Electrical and Computer Engineering Department, 
             University of California, 
             One Shields Avenue, Davis, CA 95616, USA }


\date{\today}

\begin{abstract}
An imaging system is proposed for matter-wave functions
that is based on producing a quadratic phase modulation on the wavefunction of a charged particle,
analogous to that produced by a space or time lens. The modulation is produced
by co-propagating the wavepacket within an extremum of the harmonic vector and scalar potentials 
associated with a slow-wave electromagnetic structure. By preceding and following this 
interaction with appropriate dispersion, characteristic of a solution to the time-dependent \Sch\ equation, 
a system results that is capable of magnifying ({\it i.e.}, stretching or compressing the space- and time-scales) and 
time-reversing an arbitrary quantum wavefunction.
\end{abstract}

\pacs{03.65.-w, 42.79.-e, 42.25.-p }

\maketitle 




As originally proposed by Aharanov and Bohm, the presence of a static potential
in the absence of an electric or magnetic field would be detectable owing to 
its effect on the phase of the wavefunction of a charged particle\cite{Aharonov_Bohm:59}. 
This idea was soon demonstrated\cite{Chambers:60} and the principle has spawned a
large amount of activity over the decades
 \cite{Peshkin_Tonomura:89,Hamilton:97,Batelaan_Tonomura:09}.
At the heart of the Aharanov-Bohm effect is the notion that the scalar and
vector potentials of electromagnetic theory are real, measurable, entities and
their effects distinct from the fields to which they are related. However,
an essential feature of the mechanism and proof is that the fields and 
therefore the potentials are static, for if they were not, Faraday's law and the
Amp\`ere-Maxwell relation would require the existence of electric and magnetic
fields coincident with the potentials which would contaminate the desired 
effect with forces\cite{Boyer:73}.

\begin{figure}[!ht]
\includegraphics[scale=1.04]{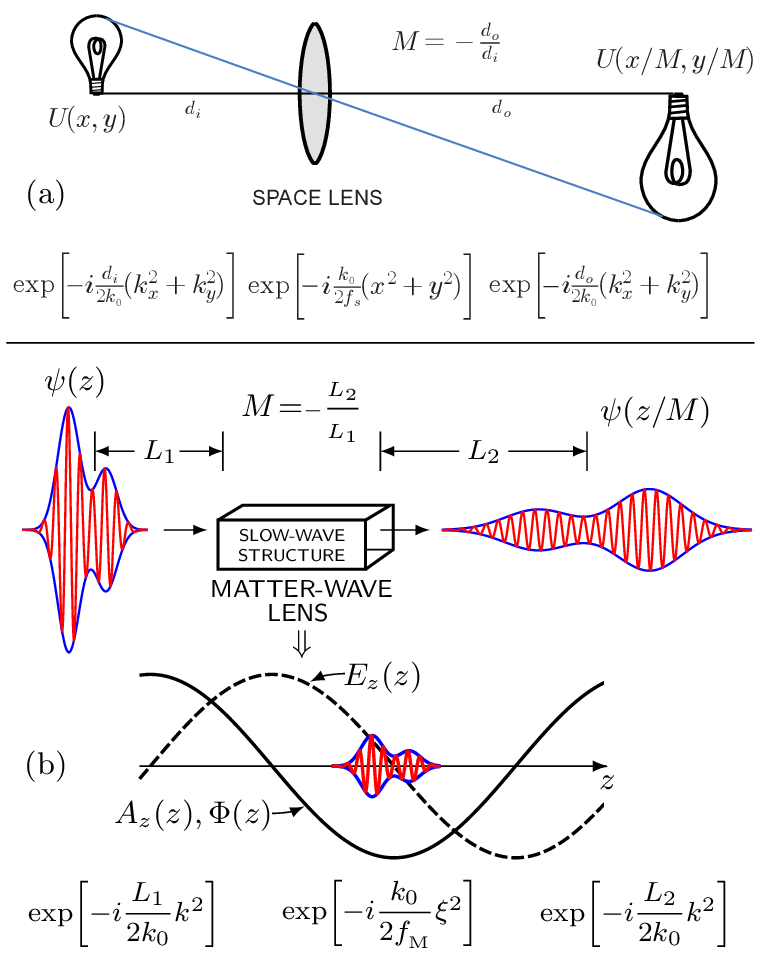}

\caption{Space-time duality between spatial and matter-wave imaging.
(a) Conventional spatial imaging system. Input and output diffraction produce 
quadratic phase filtering in Fourier spectra of the transverse coordinates.
Lens produces quadratic phase modulation directly on transverse coordinates.
(b) Imaging system for quantum wavefunctions. Input and output dispersion 
resulting from free-space propagation distances $L_1$ and $L_2$ correspond to the
object and image distances in the spatial imaging system. An electromagnetic 
slow-wave structure provides lens action by producing a quadratic phase on the dispersed
wavefunction due to the traveling-wave interaction with the scalar and vector potentials.
(Phase relationship shown for a positive charge).
The output wavefunction is a rescaled and time-reversed version of the input 
with magnification $M=-L_2/L_1$.}
\label{Matter_wave_imaging_figure} 
\end{figure}

Time-varying potentials, on the other hand, have seldom been considered in terms of their
influence on wavefunction phase\cite{Lee:92}. The purpose of this Letter
is to suggest a mechanism whereby a quadratically-varying potential (scalar
and/or vector) can produce an accumulated quadratic phase on the 
wavefunction of a charged particle in a manner consistent with the action of 
a lens. By preceding and following this interaction with the normal dispersion
inherent in wavefunction propagation, an {\sl imaging system for matter waves}
might be realized that would produce magnified and time-reversed replicas
of the original wavefunction (Fig.\ \ref{Matter_wave_imaging_figure}). 
Note that this mechanism is distinctly different from, and more general than, systems
that image the position of charged particles in space\cite{Peshkin_Tonomura:89,Hamilton:97,Batelaan_Tonomura:09} or map their evolution by time of flight\cite{Szriftgiser:96}.

The key component in the proposed system is an 
electromagnetic slow-wave structure with a longitudinal traveling-wave electric field 
that co-propagates with the wavepacket under consideration. If the
wavepacket coincides with a zero-crossing of the field 
(Fig.\ \ref{Matter_wave_imaging_figure}b) it is at an extremum
of the potential(s) and, as we will show, it will acquire a predominantly
quadratic phase shift essential for lens action, with minimal force imparted by the field. 

Central to the three 
seemingly disparate phenomena of Fresnel diffraction, narrowband dispersion and 
the quantum-mechanical description of free-particle propagation is the 
complex diffusion equation, which in general form is given by
\begin{equation}
  \frac{\partial\psi}{\partial t} =    i \alpha \frac{\partial^2 \psi}{\partial z^2} ,
\label{Sch_Eq}
\end{equation}
\noindent where the diffusivity term $\alpha$ depends on 
the particular problem at hand. 
One can consider the independent variable
$t$ as guiding the evolution of the wavefunction $\psi$ while $z$
maps its profile. In quantum mechanics the diffusivity $\alpha=\hbar/2m$
while in diffraction $\alpha=1/2k$ and in optical dispersion
$\alpha=(1/2)d^2\beta/d\omega^2$. 
(Here we use the symbols common to 
diffraction and dispersion analyses; $k=2\pi/\lambda$ is the wavenumber and $\beta=\omega n(\omega)/c$ is the phase constant.
Both are phase shifts per-unit-length.)

Based on the similarity of their 
governing equations, Fresnel diffraction and narrowband dispersion
have been successfully united in the concept of temporal imaging
which, as its name suggests, is a system for stretching or compressing
electromagnetic waveforms of carrier-envelope form while maintaining the 
integrity of the envelope profile \cite{Tournois:64,Caputi:65,Tournois:68,Akhmanov:69,Kolner:94c}.

A feature common to both diffraction and dispersion is the quadratic
phase that is introduced into the frequency spectrum of the envelope,
either of the cross-sectional profile of a beam in diffraction or the longitudinal
pulse profile in dispersion. Although quadratic in frequency, this effect
 also acquires strength linearly with the evolution variable. The same must be
true, of course, for the \Sch\ equation.

To realize imaging in both space and time, it is also necessary to produce 
quadratic phase in the real-space domain of the profile variable. 
Thus a space lens, which produces a quadratic phase transformation in the 
transverse (profile) variable has its counterpart in a time lens which produces 
quadratic phase in the  local time coordinate.\cite{Kolner:11a} To obtain
corresponding lens action for the wavefunction of a particle, we look to the 
governing p.d.e.\ \eqref{Sch_Eq} and recognize that a quadratic phase imparted
to the $z$-coordinate would suffice. As we will see, this can be accomplished 
by interaction between the particle and an electromagnetic potential, either
vector, scalar or both. 


%
The general solution to \eqref{Sch_Eq} as an initial-value problem in
one dimension, assuming a carrier-envelope type of wavepacket
and a relatively narrow
momentum spectrum centered about $k_0$,
can be written for the quantum-mechanical case as
\begin{multline}
 \psi(z,t)  = e^{i(k_0 z -\omega_0 t)}  \int_{-\infty}^\infty \psi_0 (k ,0) \\
  \times \exp \left[ -i \left( k v_g
  +  k^2 \dfrac{\hbar}{2m} \right) t \right] e^{i k z} \, \dfrac{dk}{2\pi},
\label{General_solution}
\end{multline}

\noindent where $\psi_0 (k,0)$ is the initial spectrum of the wavefunction
$\psi(z,0)$ shifted to baseband ({\it i.e.\ }$\psi_0 (k,0)=\psi(k+k_0, 0)$)
and $v_g$ is the group velocity. 

It will be especially useful in the context of this exposition to convert 
the solution \eqref{General_solution} to a traveling-wave coordinate system
moving at the group velocity of the wavepacket; $ \xi \equiv z-v_g t$
and $\tau\equiv t$. Introducing these variables transforms \eqref{General_solution}
into 
\begin{multline}
  \psi(\xi,\tau) = e^{i(k_0 \xi + \omega_0 \tau)} \int_{-\infty}^\infty \psi_0 (k ,0) \\[1ex]
 \times \exp \left( -i \dfrac{\hbar \tau}{2m}k^2  \right)  e^{i k\xi} \, \dfrac{dk}{2\pi},
\label{General_solution_TW_coordinates}
\end{multline}

\noindent which represents a stationary envelope in the $(\xi,\tau)$ reference frame
superimposed upon a backward-propagating carrier with phase velocity 
$\omega_0/k_0$. This interpretation can also be seen from the nature of the dispersion relation for 
\eqref{Sch_Eq}. For a spectrum centered at $k_0$,
\begin{equation*}
 \omega(k)= \dfrac{\hbar k^2}{2m}, \>\>
  v_g = \dfrac{d\omega}{dk}\biggr|_{k=k_0}  \hskip -2ex = \dfrac{\hbar k_0}{m}, \>
 \quad v_0=\dfrac{\omega_0}{k_0} = \dfrac{\hbar k_0}{2m},
\end{equation*}
\noindent and thus $v_g/v_0 = 2$. The quadratic phase term in 
\eqref{General_solution_TW_coordinates} will be seen 
to be an essential feature of the space-time imaging process.


To produce lens action on the quantum-mechanical wavefunction, we must
generate a quadratic phase over the dispersed envelope.
One mechanism that will accomplish this is interaction with an
electromagnetic potential. Consider a slow-wave structure designed to 
interact with charged particles which has an electric field component along the axis
of particle motion. Such structures have long been studied and developed for use in 
electron accelerators and vacuum tubes\cite{Hutter:60, Bevensee:64}.

A transverse magnetic (TM) field mode in a slow-wave guide can have an electric field
on axis that is directed exclusively along the axis of the guide. 
Assuming a monochromatic wave propagating in such a slow-wave structure
we can write the longitudinal component of the vector potential and
the scalar potential as traveling waves  
\begin{equation}
   A_z(z,t) = A_0 e^{i(k_mz-\omega_m t)},    \>\>
  \Phi(z,t) = \Phi_0 e^{i(k_m z-\omega_m t)} 
\label{Traveling_wave_potentials_1}
\end{equation}
where the subscripts ``$m$'' on the wavenumber and angular frequency indicate that
these are associated with a modulating field.
\noindent The two peak potentials can be related using the Lorentz gauge
\begin{equation}
   \nabla\cdot \mathbf{A}  = - \dfrac{1}{c^2} \dfrac{\partial \Phi}{\partial t} , 
   \hskip 5ex  \dfrac{k_m c^2}{\omega_m} A_0  = \Phi_0               
\end{equation}

\noindent so that the potentials \eqref{Traveling_wave_potentials_1} become
\begin{align}
    A_z(z,t) & = A_0 e^{i(k_m z-\omega_m t)},  
\label{Traveling_wave_potential_A}\\
   \Phi(z,t) & = \dfrac{k_m c^2}{\omega_m} A_0 e^{i(k_m z-\omega_m t)} 
\label{Traveling_wave_potential_phi}
\end{align}
From a practical standpoint, guided wave structures are generally analyzed in terms of
fields rather than potentials and thus it will be useful to relate the peak potential
$A_0$ to the peak electric field from the defining relation
(using real-valued functions)
\begin{equation*}
   \mathbf{E} (z,t) = -\nabla \Phi (z,t) - \dfrac{\partial \mathbf{A}}{\partial t} 
     = E_0 \sin (k_m z-\omega_m t) \> \zhat   \notag 
\end{equation*}
\ni and thus
\begin{equation}
  A_0 = \dfrac {E_0}{\omega_m \left(c^2/v_p^2 -1 \right) }
\label{E_0_Phi_0}
\end{equation}
where $v_p=\omega_m/k_m$ is the phase velocity of the fields and 
potentials within the slow-wave structure.

Now, assume a charged particle of mass $m$ moving at the group velocity $v_g=\hbar k_0/m$
co-propagates with the electromagnetic field in the slow-wave structure
and, furthermore, assume that it is synchronized with a peak of the potentials 
(Fig.\ \ref{Matter_wave_imaging_figure} )
where the variation is essentially quadratic.
Then, in a manner similar to the Aharonov-Bohm effect for stationary fields, 
the wavefunction will accumulate an additional phase due to the scalar (electric) and 
vector (magnetic) potentials given by
\begin{equation*}
\DelphiE  = -\dfrac{q}{\hbar}\int \Phi \, dt, \quad
\DelphiM  = \dfrac{q}{\hbar}\int \mathbf{A}\cdot d\mathbf{s} 
\end{equation*}
The combined effect of interacting with both potentials can be written as
one integral over the time coordinate using
$z^\prime = v_g t^\prime$.
Assuming an interaction length $L$ and traveling-wave potentials given by
\eqref{Traveling_wave_potential_A} and \eqref{Traveling_wave_potential_phi} 
the accumulated phase shift in the absence of dispersion becomes, using 
real-valued functions,
\begin{multline}
  \Gamma(z,L/v_g) =   \dfrac{q A_0}{\hbar} \left( v_g- \dfrac{c^2}{v_p} \right) \\
   \times \int_0^{L/v_g}  \cos (k_m z - \omega_m t + \theta)\, dt   
\label{Accumulated_phase1}
\end{multline}
\ni where  $\theta$ is an initial phase offset between the potentials and the 
wavepacket and determines whether the matter-wave lens has a positive or
negative focal length. 

Since the wavepacket is co-propagating with the modulating 
potentials, it will be useful to transform this into a traveling-wave  
coordinate system moving with the group velocity  of the wavepacket as
was done for the dispersion problem. 
Integral \eqref{Accumulated_phase1} is readily evaluated,
\begin{multline}
   \Gamma(\xi,L/v_g) = \dfrac{q A_0 L}{\hbar} \left(1- \dfrac{c^2}{v_p v_g} \right) \\[1ex]
  \times \dfrac{\sin \Delta\phi/2}{\Delta\phi/2} 
   \cos\left( k_m \xi + \Delta\phi/2 + \theta \right)  
\label{Accumulated_phase3}
\end{multline}
where
\begin{equation}
  \Delta\phi \equiv \omega_m L  \left(\dfrac{1}{v_p} - \dfrac{1}{v_g} \right)	
\end{equation}

\ni is the phase slip produced by the walkoff between the wavepacket and the 
modulating potentials. We now substitute \eqref{E_0_Phi_0} for the peak vector
potential in terms of the electric field,
\begin{multline}
   \Gamma(\xi,L/v_g) = - \dfrac{q E_0 L}{\hbar \omega_m} \dfrac{\left(c^2/v_p v_g-1 \right)}
                                                             {\left(c^2/v_p^2-1 \right)}    \\
  \times \dfrac{\sin \Delta\phi/2}{\Delta\phi/2} 
       \cos\left( k_m \xi + \Delta\phi/2 + \theta \right)  
\label{Accumulated_phase4}
\end{multline}
\ni  
Note that for the case of perfect velocity matching ({\it i.e.}, $v_p=v_g$; $\Delta\phi=0$), 
\eqref{Accumulated_phase4} simplifies considerably,
\begin{align}
   \Gamma(\xi,L/v_g) & = - \dfrac{q E_0 L}{\hbar \omega_m}  \cos\left( k_m \xi +\theta \right) \\  
                     & = - \Gamma_0  \cos\left( k_m \xi + \theta \right)  
\label{Accumulated_phase_ideal}
\end{align}
where 
\begin{equation}
 \Gamma_0\equiv q E_0 L/\hbar\omega_m 
\end{equation}
is the peak phase deviation. The initial phase offset $\theta$ is chosen
according to the sign of the charge $q$ in order that $\Gamma(\xi,L/v_g)>0$
in the vicinity of $\xi=0$, thus ensuring a positive focal length (see below),
\begin{equation}
 \theta = 
   \begin{cases}  \pi,&  q =+|q|,  \\
   	           0 ,&  q =-|q|.
   \end{cases}
\label{Phase_theta}
\end{equation}

The process of quadratic phase modulation by co-propagating the wavefunction
with a sinusoidal potential is but a specific example of any general modulation
scheme which may result in a quadratic term. As with space and time lenses, 
it will be very useful to define an equivalent
focal length and focal-length-to-aperture ratio, or $f^\#$ of a matter-wave 
lens\cite{Kolner:94d}. We can draw an analogy between the focal length of
a space lens \cite{Goodman:68}
and our matter-wave lens by comparing the phase variation with respect to the
profile variables,
\mathleft
\begin{align}
 \text{\small SPACE:} && \exp\bigl[-i\phis (x,y)\bigr]
                       & = \exp \biggl[ -i \dfrac{k}{2 \fs}(x^2 + y^2) \biggr]\notag\\
\text{\small MATTER:} &&  \exp\bigl[-i\phim (\xi)\bigr]
                       & = \exp \biggl[ -i \dfrac{k_0}{2 \fm} \xi^2 \biggr]
\label{Matter_wave_phase}
\end{align}
\mathcenter
To assign the concept
of focal length to an arbitrary phase modulation process we expand the 
phase induced by that process 
 in a Taylor series
and equate the coefficient on the second-order term to the form 
in \eqref{Matter_wave_phase}. For the matter-wave lens phase described
by \eqref{Accumulated_phase_ideal} we find
\begin{equation}  
       \fm \equiv \dfrac{k_\0}{\Gamma_\0 k_m^2}
\label{Focal_length_def}
\end{equation}

The assumption of an accumulated phase that is  generally quadratic
will be valid if the extent of the wavefunction is confined predominantly
to a maximum of the propagating cosine potential, perhaps with a 
time gate or shutter. 
We may consider this region
as defining an effective aperture for the matter-wave lens and this 
suggests the concept of $f^\#$. A reasonable estimate is approximately 
$1/2\pi$ of the period. In space, in the traveling-wave
coordinate system, this corresponds to a window $\Delta\xi=\lambda_m/2\pi=1/k_m$.
Let us then define 
\begin{equation}
  \fm^\# \equiv \dfrac{\fm}{\Delta\xi} = \dfrac{\fm}{1/k_m} = \dfrac{k_0}{\Gamma_0 k_m}
\label{F_number_matter_1}
\end{equation} 
If we now equate the classical to the quantum-mechanical momentum, $k_0=mv_g/\hbar$,
and substitute into \eqref{F_number_matter_1} along with the definition for 
$\Gamma_0$ we find
\begin{equation}
  \fm^\#  = \dfrac{m v_g v_p}{qE_0L} =  \dfrac{mc^2}{n^2 qE_0L}
\end{equation}
where, owing to the assumption of perfect velocity matching, $v_g v_p=(c/n)^2$ and
$n$ is the slowing factor for the electromagnetic 
slow wave. We see that the numerator is the rest mass energy of the 
particle, while in the denominator, $qE_0L$ would be the kinetic energy acquired by the particle
in a uniform field $E_0$ accelerated from rest 
through a distance $L$. However, in the traveling-wave frame of the particle, the electric
field is in phase quadrature with the potentials, going through a zero-crossing where the 
potentials are maximized, and thus the particle feels no net force if its center of mass
is aligned with the zero-field point.

As an example, let's assume an electron is propagating in a slow-wave structure where 
$n=10$ (corresponding to a kinetic energy of 3 keV) and the interaction length $L=1$ cm.
We find $ \fm^\#  = 5 \times 10^5/E_0 = 5$ for an electric field of 
$E_\0=10^5$ V/m (1 kV/cm). 


Imaging of matter waves is accomplished by the concatenation of dispersion,
lens action and more dispersion (Fig.\ \ref{Matter_wave_imaging_figure}).
Mathematically we see from \eqref{General_solution_TW_coordinates}
and \eqref{Matter_wave_phase} that this corresponds to quadratic phase filtering in 
wavenumber space for dispersion and quadratic phase modulation in coordinate space
for the matter-wave lens. To simplify expressing this three-step process, we introduce
the following functions

\mathleft
\begin{align}
\text{\footnotesize INPUT DISPERSION:}
&& \mathscr{G}_1 (k,\tau_1)
& = \exp\left[ -i a k^2 \right] \\
\text{\footnotesize MATTER-WAVE LENS:}
&&  H(\xi)
& = \exp \left[-i \xi^2 / 4c \right] \\
\text{\footnotesize OUTPUT DISPERSION:}
&& \mathscr{G}_2 (k,\tau_2)
& = \exp\left[ -i b k^2  \right]
\end{align}
\mathcenter
%
%
\mathleft
\begin{align}
\text{where:} \qquad   a =\dfrac{\hbar\tau_1}{2m}, \qquad   b =\dfrac{\hbar\tau_2}{2m}, 
     \qquad  c =\dfrac{\fm}{2k_\0},
\label{ABC}
\end{align}
\mathcenter
$\tau_{1,2}$ are the propagation times in the input and output 
dispersive regions, respectively, and $\fm$ is given by \eqref{Focal_length_def}.

Carrying out the forward and inverse Fourier transforms (indicated by $\mathscr{F}$
and $\mathscr{F}^{-1}$, respectively) of the three-step process of imaging
gives the wavefunction following the second (output) dispersive region,
\begin{multline}
 \psi(\xi, \tau_2)  = e^{i[k_0 \xi + \omega_0(\tau_\subone+\tau_\subtwo + \tau_\subl) + \Gamma_0 ]} \\
   \times
   \mathscr{F}^{-1} \biggl\lbrace  \mathscr{F} \biggl\lbrace  \mathscr{F}^{-1}  \biggl\lbrace
     \psi_0 (k,0) \mathscr{G}_1 (k,\tau_1) \biggr\rbrace \\
  \times  H(\xi) \biggr\rbrace
    \mathscr{G}_2 (k,\tau_2)  \biggr\rbrace
\label{Output_wavefunction_1}
\end{multline}
where $\tau_l$ is the propagation time through the matter-wave lens.
Neglecting multiplicative phase and amplitude constants, the final inverse Fourier transform
contains the essential features of the imaging problem and expresses the wavefunction 
envelope in the local traveling-wave system,
\begin{multline}
  \psi(\xi, \tau_2) \propto \int_{-\infty}^\infty \psi_0 (k,0) \\
 \exp\biggl[ i \biggl( \dfrac{1}{1/c -1/b} -a \biggr) {k}^2 
               + i \dfrac{c\,\xi}{c-b}  k \biggr] \, dk .                
\label{Output_wavefunction_3}
\end{multline}

\ni This integral represents the initial envelope spectrum $\psi_0 (k,0)$
multiplied by a quadratic spectral phase
and then inverse Fourier-transformed to a re-scaled space-time coordinate. 
In order for this wavefunction to be a replica, or ``image'', of the input waveform, we must
eliminate the quadratic phase. Thus we set $1/c -1/b = 1/a$ and find
\begin{equation}  
    \dfrac{m}{\hbar\tau_1} + \dfrac{m}{\hbar\tau_2} = \dfrac{k_\0}{\fm}
\end{equation}
or, setting $p=\hbar k_0=mv_g$ and noting that the propagation distances in the 
input and output dispersive regions are $L_1 = v_g \tau_1$ and $L_2 = v_g \tau_2$,
this becomes 
\begin{equation} 
   \dfrac{1}{L_1} + \dfrac{1}{L_2} = \dfrac{1}{\fm}	 
\label{Imaging_condition}
\end{equation}
which is the {\sl imaging condition}, familiar from classical optics.

The re-scaled space coordinate in the Fourier transform kernel takes on
an equally significant and familiar form. Substituting from
\eqref{ABC} and the imaging condition yields
\begin{equation}
  \dfrac{c-b}{c} = -\dfrac{b}{a} = -\dfrac{\tau_2}{\tau_1} = -\dfrac{L_2}{L_1} \equiv M
\label{Def_magnification}
\end{equation}
which defines the {\sl magnification.} Thus, when the imaging condition is 
satisfied, the relationship between the input and output wavefunction 
envelopes is
\begin{equation}
  \psi(\xi, \tau_2) \propto \psi(\xi/M, 0) 
\end{equation}
Notice that the magnification $M$ \eqref{Def_magnification} takes on a negative value 
and therefore produces a space- and time-reversed image of the
wavefunction. This implies no violation of causality as it is 
merely a consequence of the quadratic phase modulation and
 frequency-domain filtering of the wavefunction spectrum.
Indeed, there is no mechanism producing output prior to input. 


To this point we have not included the concept of resolution, or how fine
the structure of a wavefunction can be resolved. As in the case of conventional
optical imaging systems, limitations will arise due to the aperturing effects
at the lens. We hinted at this by assuming that the dispersed wavefunction 
entered the matter-wave lens and only interacted with $\lambda_m/2\pi$ 
of the guided electromagnetic wave in the slow-wave guide. Let's assume
we have a mechanism that creates a shutter admitting only a portion of the dispersed wavepacket 
over this duration. In concert with similar analyses in spatial and
temporal imaging systems, the impulse response of the system will be given
by the Fourier transform of this aperture function. Then, to a good approximation,
the resolution referred to the input space scale for large magnifications
can be shown to be\cite{Kolner:94c}
\begin{equation}
  \delta \xi_{in} \approx \lambda_0 \fm^{\#} 
\end{equation}
where $\lambda_0=2\pi/k_0$ is the de Broglie wavelength.

An interesting effect with time apertures that will have an impact on a stream
of particles in certain situations is the matter-wave equivalent of edge 
diffraction due to a semi-infinite opaque screen\cite{Goodman:68}. 
This was first pointed out by Moshinsky in a seminal paper in 1952\cite{Moshinsky:52} which he described as ``diffraction in time'' and is discussed, along with many other interesting transient phenomena, in the comprehensive
review by del Campo, {\sl et al.}\cite{Del_Campo:09} 
The effect of temporal slits (equivalent to the shutter discussed here) has also been studied experimentally\cite{Szriftgiser:96} and a full three-dimensional analysis of diffraction
and dispersion from a shutter has been presented by Beau and Dorlas\cite{Beau:15}.
It should be noted, however, that when
the imaging condition \eqref{Imaging_condition} is satisfied, the quadratic phase term
in \eqref{Output_wavefunction_3} is eliminated and the aperture effect is changed from
Fresnel to Fraunhofer diffraction.

The assumption that no dispersion occurs within the slow-wave structure
is not necessarily valid in all cases. There will be a trade-off between interaction length
$L$ and peak potential $A_0$ (or peak field amplitude $E_0$) to maintain a low $f$-number
and minimize dispersion. This argues for high fields and short interaction lengths
but if the latter cannot be attained, then incorporating dispersion within the slow-wave
structure can be accommodated with the definitions of the net input and output dispersions. 

Placement of the wavepacket on a cusp, or extremum, of the potential also places it 
at the zero-crossing of the electric field (Fig.\ \ref{Matter_wave_imaging_figure}). 
If the group velocity of the particle is 
not perfectly matched to the phase velocity of the potential wave then it will begin to
move and experience a nonzero force $F_z = qE_z$. However, the sign of 
the the electric field is such as to produce a {\sl restoring force} on the charged particle, 
regardless of the sign of the charge as long as \eqref{Phase_theta} is satisfied.

In summary I have proposed a system for space-time transformation and imaging of matter-wave
functions in a fashion entirely analogous to spatial or temporal imaging. It relies on the 
introduction of a quadratic phase modulation on the envelope of the wavefunction of a 
charged particle following and preceding regions of normal dispersive spreading. Recent
work in ultra-compact electron-optical accelerators \cite{Peralta:13} and coherent 
control of electrons \cite{Kfir:20,Wang:20} may make ideal candidates for 
testing some of these ideas.


\begin{acknowledgments}
This work was funded in part by the National Science Foundation under grants
ECS-9110678, ECS-9521604 and ECS-9900414, and also by the David and
Lucile Packard Foundation.
\end{acknowledgments}

\vskip 2ex
\noindent {\sffamily \bfseries\small DATA AVAILABILITY STATEMENT}

 Data sharing is not applicable to this article as no new data were created or analyzed in this study.

\end{document}